\documentclass[11pt]{article}

\def\vev#1{\langle#1\rangle}
\begin{document}

\title{Averaging Results with Theory Uncertainties\thanks{This work supported in part by the 
U.S.\ Department of Energy under grant DE-FG02-92-ER40701.}}
\author{F.~C.~Porter\\
Lauritsen Laboratory for High Energy Physics\\
California Institute of Technology\\
Pasadena, California 91125}
\date{March 27, 2008\\
Revised August 2, 2011}

\maketitle
 
\begin{center}
 {\bf ABSTRACT}
\end{center}

Combining measurements which have ``theoretical uncertainties'' is a delicate matter,
due to an unclear statistical basis. We present an algorithm based on the notion that
a theoretical uncertainty represents an estimate of bias.

\section{Introduction}

Combining measurements which have ``theoretical uncertainties'' is a delicate matter. Indeed, we are often in the position in which sufficient
information is not provided to form a completely principled combination. Here, we develop a procedure based on the interpretation
of theoretical uncertainties as estimates of bias. We compare and contrast this with a procedure that treats theoretical uncertainties along the same lines as statistical uncertainties.

Suppose we are given two measurements, with results expressed in the form:
\begin{eqnarray}
 \hat A &\pm& \sigma_A \pm t_A \nonumber\\
 \hat B &\pm& \sigma_B \pm t_B.
 \label{eqn:ABmeas}
\end{eqnarray}
Assume that $\hat A$ has been sampled from a probability distribution of the form $p_A(\hat A;\bar A,\sigma_A)$, where $\bar A$ is the mean of the
distribution and $\sigma_A$ is the standard deviation. We make the corresponding assumption for $\hat B$. The $t_A$ and $t_B$ uncertainties
in Eq.~\ref{eqn:ABmeas} are the theoretical uncertainties. We may not need to know exactly what that means here, except that the same
meaning should hold for both $t_A$ and $t_B$. We suppose that both $\hat A$ and $\hat B$ are measurements of the same quantity of physical interest, though possibly with quite different approaches.
The question is: How do we combine our two measurements?

Let the physical quantity we are trying to learn about be denoted $\theta$. Given the two results $A$ and $B$, we wish to form an 
estimator, $\hat \theta$ for $\theta$, with ``statistical'' and ``theoretical'' uncertainties expressed separately in the form:
\begin{equation}
 \hat \theta \pm \sigma \pm t.
\end{equation}
The quantities $\hat \theta$, $\sigma$, and $t$ are to be computed in terms of $\hat A, \hat B, \sigma_A, \sigma_B, t_A,$ and $t_B$.

\section{Forming the Weighted Average}

In the absence of theoretical uncertainties, we would normally combine our measurements according to the weighted average:
\begin{equation}
 \hat \theta = {{\hat A\over\sigma_A^2}+{\hat B\over\sigma_B^2}\over {1\over \sigma_A^2} + {1\over\sigma_B^2}} \pm {1\over\sqrt{{1\over \sigma_A^2} + {1\over\sigma_B^2}}}.
\end{equation}
For clarity, we are assuming for now that there is no statistical correlation between the measurements; such correlations will be incorporated later. 

In general, $\hat A$ and $\hat B$ will be biased estimators for $\theta$:
\begin{eqnarray}
 \bar A &=& \theta + b_A\nonumber \\
 \bar B &=& \theta + b_B,
\end{eqnarray}
where $b_A$ and $b_B$ are the biases. We adopt the point of view that the theoretical uncertainties $t_A$ and $t_B$ are estimates related to
the possible magnitudes of these biases. That is,
\begin{eqnarray}
 t_A &\sim& |b_A| \nonumber \\
 t_B &\sim& |b_B|.
\end{eqnarray}
We wish to have $t$ represent a similar notion.

Without yet specifying the weights, assume that we continue to form $\hat \theta$ as a weighted average of $\hat A$ and $\hat B$:
\begin{equation}
 \hat \theta = {w_A \hat A + w_B \hat B \over w_A + w_B},
\end{equation}
where $w_A$ and $w_B$ are the non-negative weights.
The statistical error on the weighted average is computed according to simple error propagation on the individual statistical errors:
\begin{equation}
 \sigma^2 = {w_A^2 \sigma_A^2 + w_B^2 \sigma_B^2 \over (w_A + w_B)^2}.
\end{equation}

The bias for $\hat \theta$ is:
\begin{equation}
 b = \langle \hat \theta - \theta\rangle = {w_A b_A + w_B b_B \over w_A + w_B}.
\end{equation}
If the theoretical uncertainties are regarded as estimates of the biases, then the theoretical uncertainty should be
evaluated with the same weighting:
\begin{equation}
 t = {w_A t_A + w_B t_B \over w_A + w_B},
 \label{eqn:weightedt}
\end{equation}
It may be noted that this posesses desirable behavior in the limit where the theoretical uncertainties are identical (completely correlated) between the two measurements:  
The theoretical uncertainty on $\hat V$ is in this case the same as $t_A=t_B$; no reduction is attained by having multiple measurements.

However, it is not quite true that the theoretical uncertainties are being regarded as estimates of bias. The $t_A$ and $t_B$ provide only estimates for the magnitudes, not the signs, of the biases. Eq.~\ref{eqn:weightedt} holds when the biases are of the same sign. If the biases are opposite sign, then we obtain
\begin{equation}
 t = {|w_A t_A - w_B t_B| \over w_A + w_B}.
 \label{eqn:weightedtos}
\end{equation}
Thus, our formula~\ref{eqn:weightedt} breaks down in some cases. For example, suppose the theoretical uncertainties are completely anticorrelated. In the case of equal weights, the combined theoretical uncertainty should be zero, because the two uncertainties are exactly canceled in the combined result. Only a statistical uncertainty remains.

Unfortunately, we don't always know whether the biases are expected to have the same sign or opposite sign. 
As a default, we adopt the procedure of Eq.~\ref{eqn:weightedt}. In the case of similar measurements, we suspect that the sign of the bias will often have the same
sign, in which case we make the right choice. In the case of quite different measurements, such as inclusive and exclusive measurements of $V_{ub}$, there is
no particular reason to favor either relative sign; we simply don't know. The adopted procedure has the property that it errs on the side of ``conservatism'' --
we will sometimes overestimate the theoretical uncertainty on the combined result.

There is still a further issue. The results of the measurements themselves can provide information on what the theoretical uncertainty could be. Consider two measurements
with negligible statistical uncertainty. Then the difference between the two measurements is the difference between the biases. If the measurements are far apart, on the
scale of the theoretical uncertainties, then this is evidence that the theoretical uncertainties are of opposite sign. We make no attempt to incorporate this information, again erring on the conservative side.

We turn to the question of choice of weights $w_A$ and $w_B$. In the limit of negligible theoretical uncertainties we want to have
\begin{eqnarray}
 w_A &=& {1\over\sigma_A^2} \\
 w_B &=& {1\over \sigma_B^2}.
 \label{eqn:simplewts}
\end{eqnarray}
Using these as the weights in the presence of theoretical uncertanties can lead to undesirable behavior. For example, suppose $t_A\gg t_B$ and $\sigma_A\ll \sigma_B$.
The central value computed with only the statistical weights ignores the theoretical uncertainty. A measurement with small theoretical uncertainty may
be given little weight compared to a measurement with very large theoretical uncertainty. While not ``wrong'', this does not make optimal use of the
available information. We may invent a weighting scheme which incorporates both
 the statistical and theoretical uncertainties, for example combining them in quadrature:
\begin{eqnarray}
 w_A^\prime &=& {1\over\sigma_A^2 + t_A^2} \nonumber \\
 w_B^\prime &=& {1\over \sigma_B^2 + t_B^2}.
\label{Eqn:weights}
\end{eqnarray}

Any such scheme can lead to unattractive dependence on the way measurements may be associatively combined. In order to have associativity in combining three measurements $A,B,C$, we must have that the weight for the combination of any two to be equal to the sum of the weights for those two, e.g., $w_{AB}=w_A+w_B$. This is inconsistent with our other requirements. We shall adopt the procedure in Eq.~\ref{Eqn:weights}, with the understanding that it is best to go back to the original measurements when combining several results, rather than making
successive combinations.

\section{Inconsistent Inputs}

It may happen that our measurements are far enough apart that they appear inconsistent in terms of the quoted uncertainties.
Our primary goal may be to test consistency between available data and a model, including whatever theoretical uncertainties exist in the comparison. We prefer to avoid making erroneous claims of inconsistency, even at the cost of some 
statistical power. Thus, we presume that when two measurements of what is assumed to be the same quantity appear inconsistent, something is
wrong with the measurement or with the thoeretical uncertainties in the computation. If we have no good way to determine in detail where the fault lies, we adopt a method similar to that used by the Particle Data Group (PDG)\cite{bib:PDG} to enlarge the stated uncertainties. 

Given our two measurements as discussed above, we define the quantity:
\begin{equation}
 \chi^2 \equiv w_A(\hat A-\hat\theta)^2 + w_B(\hat B -\hat\theta)^2.
\end{equation}
In the limit of purely statistical and normal errors, this quantity is distributed according to a chi-square with one degree of freedom.
In the more general situation here, we don't know the detailed properties, but we nonetheless use it as a measure of the consistency
of the results, in the belief that the procedure we adopt will still tend to err toward conservatism.

If $\chi^2\leq 1$, the measurements are deemed consistent. On the other hand, if $\chi^2>1$, we call the measurements inconsistent, and apply a scale factor to the errors in order to obtain $\chi^2=1$. We take the point of view that we don't know which measurement (or both) is flawed, or whether the problem is with the statistical or theoretical error evaluation. If we did have such relevant information, we
could use that in a more informed procedure. Thus, we scale all of the errors ($\sigma_A$, $\sigma_B$, $t_A$, $t_B$) by a factor:
\begin{equation}
 S  = \sqrt{\chi^2}.
\end{equation}
This scaling does not change the central value of the averaged result, but does scale the statistical and theoretical uncertainties by the
same factor.

\section{Relative Errors}

We often are faced with the situation in which the uncertainties are relative, rather than absolute. In this case, the model
in which $\theta$ is a location parameter of a Gaussian distribution breaks down. However, it may be a reasonable approximation
to continue to think in terms this model, with some modification to mitigate bias. We also continue to work in the context of a
least-squares minimization, although it might be interesting to investigate a maximum likelihood approach.

Thus, suppose we have additional experimental uncertainties $s_A$ and $s_B$, which scale with $\theta$:
\begin{eqnarray}
 s_A &=& r_A\theta, \nonumber \\
 s_B &=& r_B\theta.
\end{eqnarray}
If $s_k$ is what we are given, we infer the proportionality constants according to $r_A=s_A/\hat A$ and $r_B=s_B/\hat B$.

The weights that are given in Eqn.~\ref{Eqn:weights} are modified to incorporate this new source of uncertainty according to:
\begin{eqnarray}
 w_A^\prime &=& {1\over\sigma_A^2 + (r_A\hat\theta)^2 + t_A^2} \nonumber \\
 w_B^\prime &=& {1\over \sigma_B^2 + (r_B\hat\theta)^2 + t_B^2}.
\label{eqn:weightsWithRelativeErrors}
\end{eqnarray}
Note that, as we don't know $\theta$, we use $\hat\theta$ instead. This means that the averaging process is now
iterative, until convergence to a particular value of $\hat\theta$ is obtained.

Likewise, there may be a theoretical uncertainty which scales with $\theta$, and we may treat this similarly. Thus, suppose that, for example,
$t_A^2 = t_{aA}^2 + t_{rA}^2$, where $t_{aA}$ is an absolute uncertainty, and $t_{rA} = \rho_A\theta$. We simply replace $\theta$ by $\hat\theta$ and substitute this expression wherever $t_A$ appears, e.g., in Eqn.~\ref{eqn:weightsWithRelativeErrors}. That is:
\begin{equation}
 w_A^\prime = {1\over\sigma_A^2 + (r_A\hat\theta)^2 + t_{aA}^2 + \rho_A^2\hat\theta^2}. 
\label{eqn:weightsWithRelativeErrorsT}
\end{equation}

\section{Summary of Algorithm}

We summarize the proposed algorithm, now including possible statistical correlations: 
Suppose we have $n$ measurements $\{x_i | i=1,2,\ldots, n\}$ with
covariance matrix
\begin{equation}
 M_{ij} \equiv \langle (x_i-\vev{x_i})(x_j-\vev{x_j})\rangle,
\end{equation}
and mean values
\begin{equation}
 \vev{x_i} = \theta + b_i.  
\end{equation}
Note that, in the non-correlated case, $M_{ij}=\sigma_i^2 \delta_{ij}$, or including relative uncertainties,
$M_{ij} = \delta_{ij}(\sigma_i^2+r_i^2\vev{x_i}^2)$.
The parameter we are trying to learn about is $\theta$, and the $b_i$ is the bias that is being estimated with theoretical uncertainties $t_i$.

The present notion of the weighted average is that we find a $\theta$ which minimizes:
\begin{equation}
 \chi^2 = \sum_{i,j}\left(x_i-\theta\right) W_{ij} \left(x_j-\theta\right).
\end{equation}
This is based on the premise that we don't actually know what the biases are, and we do the
minimization with zero bias in the $(x-\theta)$ dependence. The possible size of bias is taken into
account in the weighting, giving more weight to those measurements in which the size of the bias is likely to be
smaller.

The ``weight matrix'' $W$ in principle could be taken to be:
\begin{equation}
 (W^{-1})_{ij} = M_{ij} + t_it_j.
 \label{Eqn:weightMatrixTry}
\end{equation}
That is, $W^{-1}$ is an estimate for
\begin{equation}
 \langle (x_i-\theta)(x_j-\theta)\rangle = M_{ij} + b_ib_j.
\end{equation}
However, we don't assume that we know the relative signs of $b_i$ and $b_j$. Hence, the off-diagonal $t_it_j$ term in Eqn.~\ref{Eqn:weightMatrixTry}
could just as likely enter
with a minus sign. We therefore use the weight matrix:
\begin{equation}
 (W^{-1})_{ij} = M_{ij} + t_i^2\delta_{ij}.
\end{equation}
If we do know the relative signs of the biases, for example because the theoretical uncertainties are correlated, then 
the off-diagonal terms in Eqn.~\ref{Eqn:weightMatrixTry} should be included, with the appropriate sign. If there are 
several contributions to the theoretical uncertainty, then each piece can be treated separately, if enough is
known.
When using the term ``correlated'' with theoretical uncertainties, it should be kept in mind that it does
not necessarily have a statistical interpretation.

Setting $d\chi^2/d\theta\big|_{\theta=\hat\theta} = 0$ gives
the central value (``best'' estimate):
\begin{equation}
 \hat\theta = {\sum_{i,j} W_{ij} x_j \over \sum_{i,j} W_{ij} }.
\end{equation}
The statistical uncertainty is
\begin{equation}
 \sigma = {\sqrt{\sum_{i,j} \left(WMW\right)_{ij}} \over \sum_{i,j} W_{ij}}
\end{equation}
Note that this reduces to
\begin{equation}
 \sigma = {1\over \sqrt{\sum_{i,j} (M^{-1})_{ij} }},
\end{equation}
in the case of only statistical uncertainites.
The theoretical uncertainty is
\begin{equation}
 t = {\sum_{i,j} W_{ij} t_j \over \sum_{i,j} W_{ij} },
\end{equation}
where
\begin{equation}
 t_j \equiv \sqrt{t_{aj}^2 + (\rho_j\hat\theta)^2}.
\end{equation}

Finally, if $\chi^2 > n-1$, these error estimates are scaled by a factor:
\begin{equation}
 S  = \sqrt{{\chi^2 \over n - 1}},
\end{equation}
where $\chi^2$ here is the value after the minimization.

\section{Comparison with treating theoretical uncertainties on same footing as statistical}

Another approach to the present problem is to simply treat the theoretical uncertainties as if
they were statistical.\cite{bib:HFAG} This procedure gives the same estimator as above for $\theta$,
as well as for the statistical uncertainty $\sigma$. 
However, the results for the theoretical uncertainty differs in general.

Let $\sigma^\prime$ be the estimated statistical uncertainty on the average for this approach, and let
$t^\prime$ be the estimated theoretical uncertainty. Also, let $T_{ij}$ be the ``covariance matrix''
for the theoretical uncertainties in this picture. Then the statistical and theoretical uncertainties
on the average are given by:
\begin{eqnarray}
 \sigma^{\prime} &=& {\sqrt{\sum_{i,j} \left(WMW\right)_{ij}} \over \sum_{i,j} W_{ij}} = \sigma, \\
 t^\prime &=& {\sqrt{\sum_{i,j} \left(WTW\right)_{ij}} \over \sum_{i,j} W_{ij}}.
\end{eqnarray}
Note that the weights are given, as before, by
\begin{equation}
 W_{ij} = (M+T)^{-1}_{ij}.
\end{equation}
That is, the weights are the same as the treatment earlier, if the same assumptions about theoretical 
correlations are made in both places.

The estimated theoretical uncertainties differ between the two methods, since, in 
general, $t^\prime\ne t$. We can obtain some insight into the difference by supposing that
the matrix $T$ is diagonal: $T_{ij} = \delta_{ij}t_i^2$. Thus, 
\begin{equation}
t^\prime = {\sqrt{\sum_k t_k^2\sum_{i,j} W_{ik}W_{jk}}\over \sum_{i,j} W_{ij}}.
\end{equation}
The difference between $t$ and $t^\prime$ is that $t$ is computed as a weighted average of the individual $t$'s,
with weight $\sum_i W_{ik}$ for $t_k$, while
$t^\prime$ is a weighted average of the $t^2$'s, with weight $\sum_{i,j} W_{ik}W_{jk}$ for $t_k^2$.
Again, the approach for $t$ is based on the
notion that the theoretical uncertainties are estimates of bias, but with a conservative treatment of
any unknown correlations; the $t^\prime$ approach may be appropriate if
the theoretical uncertainties are given a probablistic interpretation.

Let's consider some possible special cases. Suppose that all of the $t_i$'s are the same, equal to $t_1$, and suppose that
the theory uncertainties are presumed to be ``uncorrelated''. In this case,
\begin{eqnarray}
 t &=& t_1 \\
 t^\prime &=& t_1{\sqrt{\sum_{i,j}(W^2)_{ij}}\over\sum W_{ij}}.
\end{eqnarray}
In the simple case where the weight matrix is a multiple of the identity, $W=wI$, this gives
$t^\prime= t_1/\sqrt{n}$.
Which is more reasonable? That depends on how we view the meaning of ``uncorrelated'' in our assumption, and on whether we
assign a probabilistic interpretation to the theoretical uncertainties. If we are supposing that
the acutal theoretical uncertainties are somehow randomly distributed in sign and magnitude, then it is reasonable to expect that
the result will become more reliable as more numbers are averaged. However, if we consider the theoretical uncertainties as
estimates of bias, which could in fact all have the same sign, then the weighted linear average is plausible. It is at least a 
more conservative approach in the absence of further information on the real nature and correlations of the uncertainties. 

Note that if the correlation in theoretical uncertainty is actually known, the weighted linear average will take that
into account. For example, suppose there are just two measurements, with $t_2=-t_1$ [and $T=t_1^2\pmatrix{1&-1\cr -1&1\cr}$]. If the weights are the same [that is, we also
have $\sigma_1=\sigma_2$, or rather, we have $M=\sigma_1^2\pmatrix{1&\rho\cr \rho&1\cr}$] then $t=0$. The other approach likewise gives $t^\prime=0$. 

A different illustrative case is when $t_2=0$, $t_1\ne 0$, and $M=\pmatrix{\sigma^2_1&0\cr 0&\sigma_2^2\cr}$. In this case, we find
\begin{eqnarray}
 \hat\theta &=& \hat\theta^\prime = \left({x_1\over \sigma_1^2+t_1^2} + {x_2\over \sigma_2^2}\right)\bigg/\left({1\over \sigma_1^2+t_1^2} + {1\over \sigma_2^2}\right),\\
 \sigma &=& \sqrt{{\sigma_1^2\over (\sigma_1^2+t_1^2)^2} + {1\over \sigma_2^2}}\bigg/\left({1\over \sigma_1^2+t_1^2} + {1\over \sigma_2^2}\right),\\
 t &=& t_1{{1\over \sigma_1^2+t_1^2} \over {1\over \sigma_1^2+t_1^2} + {1\over\sigma_2^2}},\\
 \sigma^\prime &=& \sigma,\\
  t^\prime &=& t.
\end{eqnarray}
Thus, when there is only one theory uncertainty, both methods get the same result (even if
correlations are present in $M$) .

\section{Conclusions}

We suggest a procedure to treat theoretical uncertainties when combining measurements of
a quantity. It must be emphasized that, at least without more information about the nature of the
theoretical uncertainty, there is no rigorous procedure, e.g., in the context of frequencies. The
interpretation we adopt is that theoretical errors represent estimates of bias. This leads to 
a straightforward algorithm. If the sign of the bias is not specified (the usual situation), the
procedure is designed to be ``conservative'', in the sense that we may err on the side of overstating
theory uncertainties on the combined result. There is also some arbitrariness in construction of the weights from the statistical and theoretical uncertainties; we suggest simply adding the uncertainties in quadrature.

Our procedure is compared with a method that treats theoretical uncertainties as if they are of statistical
origin. While there are some reassuring similarities, there are also differences. For example, the two procedures lead to different scaling behavior for the theoretical uncertainty with the number of measurements. Depending on interpretation, either result could be regarded as reasonable; our procedure yields a more ``conservative'' scaling.  In the case where there are no theoretical uncertainties, both procedures yield identical, conventional results.
It may be useful in practice to compute the uncertainties via both approaches, giving an idea for how sensitive the result is
to the assumptions.

It would, of course, be nice to have a test of the procedure. For example, does it lead to 
appropriate frequency coverage? Given the lack of clarity (and perhaps lack of consistency) in what theoretical uncertainties really mean, a meaningful test is difficult. We thus rely on the conservative nature of
our procedure -- the intervals obtained for the combined results will ``probably'' over-cover if theoretical uncertainties are present.

\begin{center}
{\bf ACKNOWLEDGMENTS
}
\end{center}

I am grateful to Gregory Dubois-Felsmann, Gerald Eigen, David Hitlin, and Robert Kowalewski for stimulating discussions on this topic.

\end{document}